\documentclass[10pt]{iopart}
\usepackage[utf8]{inputenc}
\usepackage{iopams}
\usepackage{graphicx}
\usepackage{cite}
\usepackage[pdfstartview=FitH, pdfpagemode=None, pdfpagelayout=OneColumn, colorlinks=true, linkcolor=blue, citecolor = blue]{hyperref}
\newcommand{\text}[1]{\mathrm{#1}}

\begin{document}
\title{Evolution of equilibrium particle beams under external wakefields}
  \author{M.A. Baistrukov}
\address{Novosibirsk State University, Novosibirsk, 630090, Russia}
\ead{M.A.Baistrukov@inp.nsk.su}
\author{K.V. Lotov}
\address{Novosibirsk State University, Novosibirsk, 630090, Russia}
\ead{K.V.Lotov@inp.nsk.su}

\vspace{10pt}
\begin{indented}
\item[]\today
\end{indented}

\begin{abstract}
A beam of ultrarelativistic charged particles in a plasma can reach equilibrium with its own radial wakefield and then propagate with little change in shape.
If some co-moving perturbation appears ahead of the beam, it may or may not destroy the beam with its wakefield, depending on the phase and amplitude of the wakefield.
We numerically study which perturbations can destroy a single short bunch or a train of many short bunches at the parameters of interest for plasma wakefield acceleration in an axysimmetric configuration, and how fast.
We find that there are particularly dangerous wakefield phases in which the beam can be destroyed by perturbations of very low amplitude. We also find that perturbations with an amplitude larger than the wakefield of a single bunch in the train are always destructive.
\end{abstract}

\vspace{2pc}

\ioptwocol
 \section{Introduction}

The propagation of tightly focused, short beams of high-energy charged particles in plasmas is actively studied in the context of advanced acceleration techniques \cite{NJP23-031101}. These beams can collectively interact with plasmas, either driving strong electric fields (called `wakefields') or being accelerated or focused by these fields. In many cases, the beam density is much lower than or comparable to the plasma density. This interaction regime is called `linear' \cite{PF30-252} or `weakly nonlinear' \cite{PRA39-1586,PoP13-113102}, as opposed to `highly nonlinear' or `blowout' regimes \cite{PRA44-6189,PRE64-045501}.

In the linear or weakly nonlinear regime, the particle beam entering the plasma quickly reaches a radial equilibrium with the wakefield, much faster than the energy exchange with the plasma occurs \cite{PFB2-1376,NIMA-410-461}. The equilibrium state is rather exotic: the beam density is strongly peaked near the axis, and the particle distribution in the momentum space is not Gaussian~\cite{PoP24-023119}.

We consider the drive beams that create the wakefield.
Ideally, an equilibrium drive beam propagates in an unperturbed plasma, preserving its shape, and only the particle energy reduces.
However, there may be conditions in which some perturbation (another beam or a laser pulse) appears ahead of the beam after it has reached the radial equilibrium.
The perturbation wakefield has a phase velocity equal to the speed of light $c$, so it is stationary in the beam frame and can strongly modify the beam shape, even if the wave amplitude is small. Such situations can occur in various plasma-based acceleration schemes.
For example, if the beam is manually composed of several bunches (a so-called `comb' beam \cite{NIMA-292-12, AIP569-605, IJMPB21-415, PRL101-054801, PRST-AB13-052803, NIMB-309-183, PoP27-033105}) and the first bunch has a smaller charge than the others, this bunch reaches equilibrium later than the others, and its wakefield acts on them as an external perturbation.
In some experiments planned at the AWAKE facility \cite{NIMA-829-3, NIMA-829-76, PPCF60-014046}, the bunch train formed by self-modulation of a long proton beam in the first plasma section \cite{PRL104-255003, PoP22-103110, PRL122-054801, PRL122-054802} enters the second section, where it can be perturbed either by an externally injected electron bunch \cite{JPCS1596-012008, IPAC21-Patric}, or by the proton beam head that passed through the first section without modulation \cite{IPAC16-2557, IPAC21-Patric}.
The latter is possible if the plasma in two sections is created differently: in the first section, a laser pulse propagates together with the proton beam and ionizes a rubidium vapor \cite{NIMA-740-197, JPD51-025203} so that the leading part of the proton beam (the `beam head') propagates in the neutral gas and does not self-modulate; and the second section is ionized in advance \cite{JPCS1596-012008, PPCF60-075005, IPAC14-1470}.
Hybrid acceleration schemes \cite{PRL104-195002,ApplSci9-2626} can produce beams in which one electron bunch is affected by the wakefield of another, preceding bunch \cite{PoP17-123104, NIMA-829-422, ApplSci9-2561, PoP21-123113, PRX10-041015}.
In addition, uncontrolled disturbances caused by dark currents \cite{PPCF58-034009, PRAB19-101303, PTRSA377-20180184} can occur.

All of the above shows that it is important to know the resilience of equilibrium beams to external perturbations.
In this paper, we study how an external wakefield can change a particle bunch or bunch train, what perturbation amplitude is dangerous, and how long does it take to destroy the beam.
To present the results in a general form, we use dimensionless units related to beam and plasma parameters.
In section~\ref{s2-methods}, we describe the methods of study and introduce quantitative characteristics of beam destruction.
In section~\ref{s3-effect}, we consider the changes in a single bunch and a train of 10 bunches caused by switching the perturbation on or off and analyze why perturbations with certain phases are especially dangerous for the beam.
We then discuss the time scale of beam changes in section~\ref{s4-time} and summarize the main findings in section~\ref{s5-summary}.

\begin{figure}
\includegraphics[width=0.95\columnwidth]{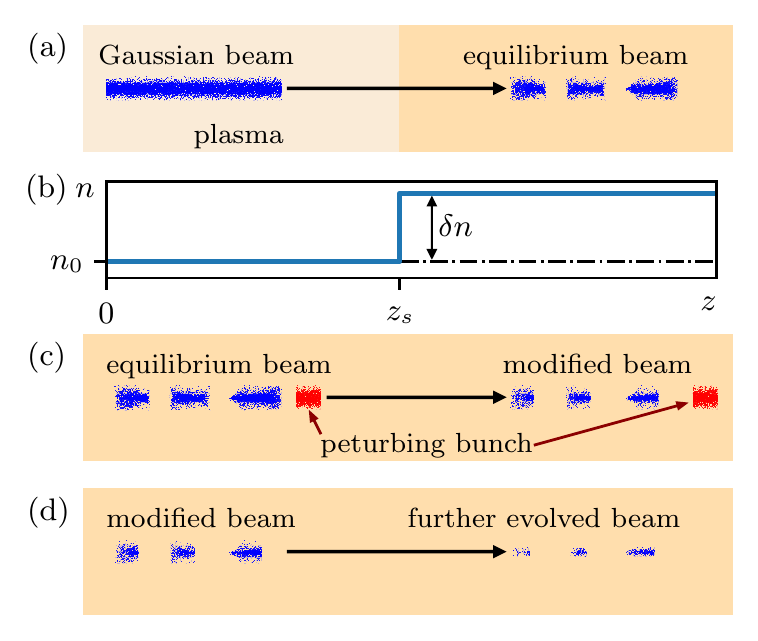}
\caption{Simulated stages of beam evolution: (a) a Gaussian beam reaches a radial equilibrium, propagating in a plasma with a density step schematically shown in (b), (c) the equilibrium beam is distorted by the wakefield of the perturbing bunch, and (d) the distorted beam evolves to another equilibrium state after the perturbation is removed.}
\label{fig1-scheme}
\end{figure}

\section{Methods}
\label{s2-methods}

The equilibrium state of a beam in plasma is difficult even to characterize analytically \cite{PoP24-023119}, let alone analytically examine its response to perturbations.
So the only way to study beam resilience is through numerical simulations. For this, we use the quasistatic axisymmetric code LCODE \cite{PRST-AB6-061301,NIMA-829-350}.
To create an equilibrium beam, we inject a beam with a Gaussian radial profile into the plasma and simulate its propagation until its shape stabilizes.
This brings our study closer to experimentally realizable conditions in which an equilibrium beam cannot be pre-formed outside the plasma.
To produce a train of several bunches, we inject a long constant-current beam into the plasma and allow the seeded self-modulation to cut the beam into equilibrium micro-bunches (figure~\ref{fig1-scheme}(a)).
The longitudinal plasma density profile $n(z)$ in this case must have a small density step-up at some distance from the entrance (figure~\ref{fig1-scheme}(b)).
Otherwise, in a constant-density plasma, the beam loses too much charge, transforming into a bunch train \cite{PoP18-024501,PoP22-103110}.

Once the equilibrium beam is formed, we introduce an external perturbation by adding a short non-evolving perfectly aligned particle bunch moving ahead of the beam (figure~\ref{fig1-scheme}(c)).
The radial profile of the disturbing beam is the same as that of the original Gaussian beam.
This narrows the generality of our study, but to the most interesting special case, in which the perturbation has the same radial spatial scale as the beam.

It is possible that if the perturbation acts during a limited time, the beam adapts to the perturbation, but degrades after the perturbation is turned off.
We analyze this option by removing the perturbation after some distance (figure~\ref{fig1-scheme}(d)).

\begin{table}[tb]
\caption{Parameters of the study.}\label{t1}
\small
\begin{tabular}{ll}\hline
  Parameter and notation & Value \\ \hline
  \textit{Beam:} &\\
  Peak density, $n_{b0}$ & $4\times 10^{-3} n_0$ \\
  Radius, $\sigma_r$ & $0.5 c/\omega_p$ \\
  Relativistic factor, $\gamma_b$ & 1000 \\
  Angular spread & $2\times 10^{-4}$ \\
  Energy spread & 0\,\% \\
  \textit{Plasma:} &\\
  Density step, $\delta n$ & $0.085 n_0$ \\
  Location of the density step, $z_s$ & $360 c/\omega_p$ \\
  \textit{Simulations:} & \\
  Domain size in $r$ and $\xi$ & $10 c/\omega_p$ and $80 c/\omega_p$ \\
  Grid size in both $r$ and $\xi$ & $0.01 c/\omega_p$ \\
  Time step for the beam & $10 \omega_p^{-1}$ \\
  Perturbation switched on at, $z_0$ & $10^4 c/\omega_p$ \\
  Perturbation switched off at, $z_1$ & $3 \times 10^4 c/\omega_p$ \\
  Final check of beam state at, $z_2$ & $5 \times 10^4 c/\omega_p$ \\
  \hline
\end{tabular}
\end{table}

In simulations, we measure times in units of $\omega_p^{-1}$ and distances in units of $c/\omega_p$, where $\omega_p = \sqrt{4 \pi n_0 e^2/m}$ is the plasma frequency, $n_0$ is the initial plasma density, $m$ is the electron mass, and $e$ is the elementary charge.
The initial beam parameters (table~\ref{t1}) are the same as in~\cite{PoP22-103110}, because we know the plasma density profile that provides efficient micro-bunching for this beam.
This parameter set does not correspond to any experiment and is chosen to study the beam perturbation by an external wakefield in its purest form and exclude complicating effects.
The beam density is low enough that the plasma response remains linear, the small angular spread excludes the emittance-driven beam divergence \cite{PoP22-123107} from consideration, and the relativistic factor is sufficiently high to distinguish the timescales of radial dynamics and beam depletion \cite{NIMA-410-461}.
The plasma is radially uniform, and the plasma ions are immobile.
All processes in the linearly responding plasma are symmetric with respect to the sign of the beam charge, so we simulate a positron beam.
We use cylindrical coordinates $(r, \varphi, z)$ and the co-moving coordinate $\xi = z - ct$.
The initial beam density at the entrance to the plasma (at $z=0$) is
\begin{equation}\label{e1}
    n_b = \cases{
        n_{b0} e^{-r^2/(2 \sigma_r^2)},\quad & $\xi<0$, \\
        0, & $\xi\geq 0$,
    }
\end{equation}
so the self-modulation is seeded by the steep leading edge of the beam.
We consider either the first bunch formed as a result of self-modulation or the train of the first ten bunches.
The simulation parameters are given in table~\ref{t1}.

The timescale for the transverse beam dynamics is
\begin{equation}\label{e2}
    \tau_0 = \sqrt{ \frac{m_b \gamma_b}{2 \pi n_{b0} e^2} },
\end{equation}
where $m_b$ is the mass of beam particles.
In our case, $\tau_0 \approx 700 \omega_p^{-1}$. This is also the typical growth time for the self-modulation \cite{PRL107-145003,PRL107-145002}, so we take $\tau_0$ as the natural unit of time.

The wave strength is conveniently characterized by the wakefield potential $\Phi$, the gradient of which describes the electromagnetic force acting on the beam particles:
\begin{equation}\label{e3}
    -\frac{\partial \Phi}{\partial \xi} = E_z, \quad
    - \frac{\partial \Phi}{\partial r} = E_r - B_\varphi,
\end{equation}
where $\vec{E}$ and $\vec{B}$ are the electric and magnetic fields.
In a linearly responding plasma, the wakefield potential on the axis is related to the beam density $n_b$ as \cite{PAcc20-171}
\begin{equation}\label{e4}
    \Phi(\xi,z) = -k_p \int_\xi^\infty \sin \bigl( k_p(\xi'-\xi) \bigr) I_\text{eff} (\xi',z) \, d\xi',
\end{equation}
where
\begin{equation}\label{e5}
    k_p = \cases{
        \omega_p/c,\quad & $z < z_s$, \\
        \omega_p \sqrt{n / n_0}\bigr / c, & $z \geq z_s$
    }
\end{equation}
is the local wavenumber determined by the local plasma density $n$,
\begin{equation}\label{e6}
    I_\text{eff} (\xi,z) = 4 \pi e \int_0^\infty K_0(k_p r) n_b(r, \xi, z) \, r dr
\end{equation}
is the effective current that we introduce to conveniently characterize the contributions of beam parts, and $K_0$ is the modified Bessel function.
The dimension of the effective current (\ref{e6}) is intentionally chosen equal to the dimension of the potential, rather than the current, to reflect their close relationship.

The leading edge of the beam (\ref{e1}) drives the wakefield
\begin{equation}\label{e7}
    \Phi (\xi, 0) = \Phi_0 \bigl( \cos (\omega_p \xi/c) - 1 \bigr),
\end{equation}
where
\begin{equation}\label{e8}
    \Phi_0 = 4 \pi e n_{b0} \int_0^\infty K_0(\omega_p r/c) e^{-r^2/(2 \sigma_r^2)} \, r dr
\end{equation}
is the effective current of the initial beam and, at the same time, the amplitude of wakefield oscillations driven by the steep beam front.
We take $\Phi_0$ as the unit of both wakefield strength and effective current.

\begin{figure}[tb]
\includegraphics{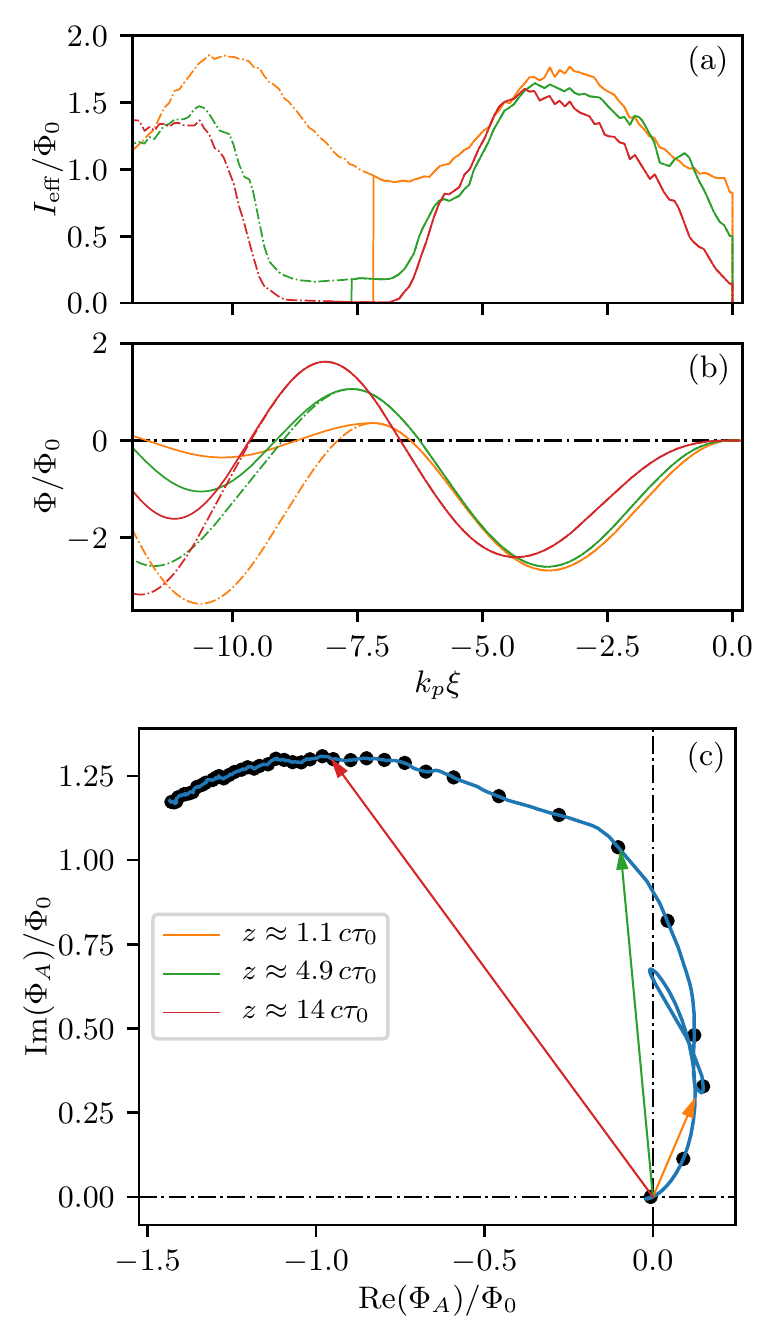}
\caption{Transition from the constant-current beam to equilibrium bunches: (a) effective current $I_\text{eff} (\xi)$ and (b) wakefield potential on the axis $\Phi(\xi)$ at different propagation distances $z$ (shown in different colors), and (c) the complex wakefield amplitude $\Phi_A (z)$ after the first bunch.
The solid lines in (a) and (b) show the current and field of the first bunch only, and the dashed lines are for the bunch train.
Black points in (c) follow at intervals of $c\tau_0$ and show the time scale of the process.
The colored arrows in (c) correspond to the cases shown in (a) and (b).}
\label{fig2-equilibration}
\end{figure}

We characterize the evolution of individual bunches and the beam as a whole by the change of the wakefield they excite.
Changes in both amplitude and phase of the wakefield are important, so we introduce the complex wakefield potential $\widetilde{\Phi}$ and its complex amplitude $\Phi_A$, which contain information about both:
\begin{equation}\label{e9}
    \widetilde{\Phi}(\xi,z) = \Phi_A (\xi,z) e^{i k_p \xi},
\end{equation}
\begin{equation}\label{e10}
    \Phi_A (\xi,z) = -i k_p \int_\xi^\infty I_\text{eff} (\xi',z) e^{-i k_p \xi'} \, d\xi',
\end{equation}
\begin{equation}\label{e11}
    \Phi (\xi,z) = \text{Re} \bigl( \widetilde{\Phi} \bigr), \quad
    E_z(r=0,\xi,z) = k_p \text{Im} \bigl( \widetilde{\Phi} \bigr).
\end{equation}
Similarly to~\cite{PoP22-103110}, we define the boundary between the bunches as the cross-section in which the wakefield potential has a local maximum and therefore quickly defocuses the particles.
This definition allows us to unambiguously identify the bunches, even if the gaps between them have not yet formed.

\begin{figure}[tb]
\includegraphics{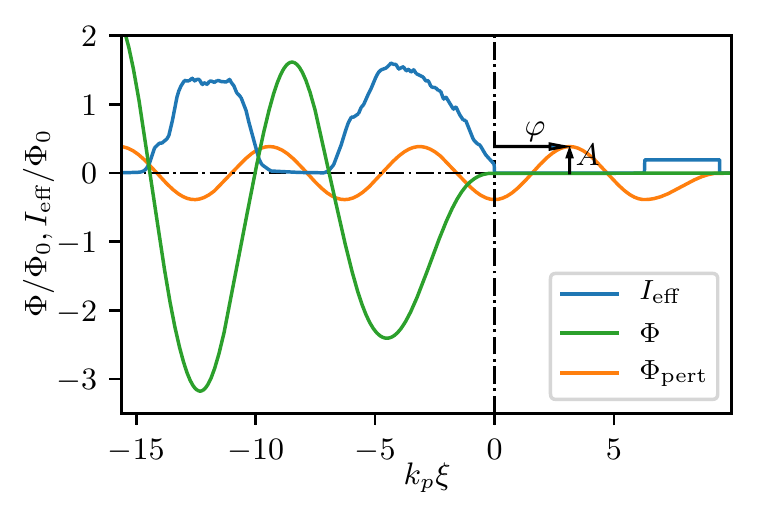}
\caption{The effective current $I_\text{eff}$ of the perturbing bunch and first two equilibrium bunches, the wakefield potential $\Phi$ of the equilibrium beam, and the wakefield potential $\Phi_\text{pert} = A e^{-i\varphi}$ of the perturbing bunch.}
\label{fig3-perturbation}
\end{figure}

During the self-modulation, the shape and the effective current of the bunches first change quickly and then stabilize and change much slower.
The complex amplitude behaves similarly (figure~\ref{fig2-equilibration}).
The evolution of the beam never stops completely because of the gradual depletion of its energy and the emittance-driven erosion of the beam head.
Therefore, there is some freedom in choosing the state that we take as the equilibrium.
For certainty, we consider the beam at $z = z_0 \approx 14 c \tau_0$ as the equilibrium one (red lines in figure~\ref{fig2-equilibration}).
After the beam passes this distance, we introduce a perturbing positron bunch, which creates a wakefield of amplitude $A$ and relative phase $\varphi$ (figure~\ref{fig3-perturbation}).
The perturbing bunch has the length $\pi k_p^{-1}$ and the same radial density profile (\ref{e1}) as the initial beam.
It propagates exactly at the speed of light and does not change shape.
By varying the longitudinal position and density of this bunch, we control the amplitude and phase of the perturbation.

\begin{figure}[tb]
\includegraphics{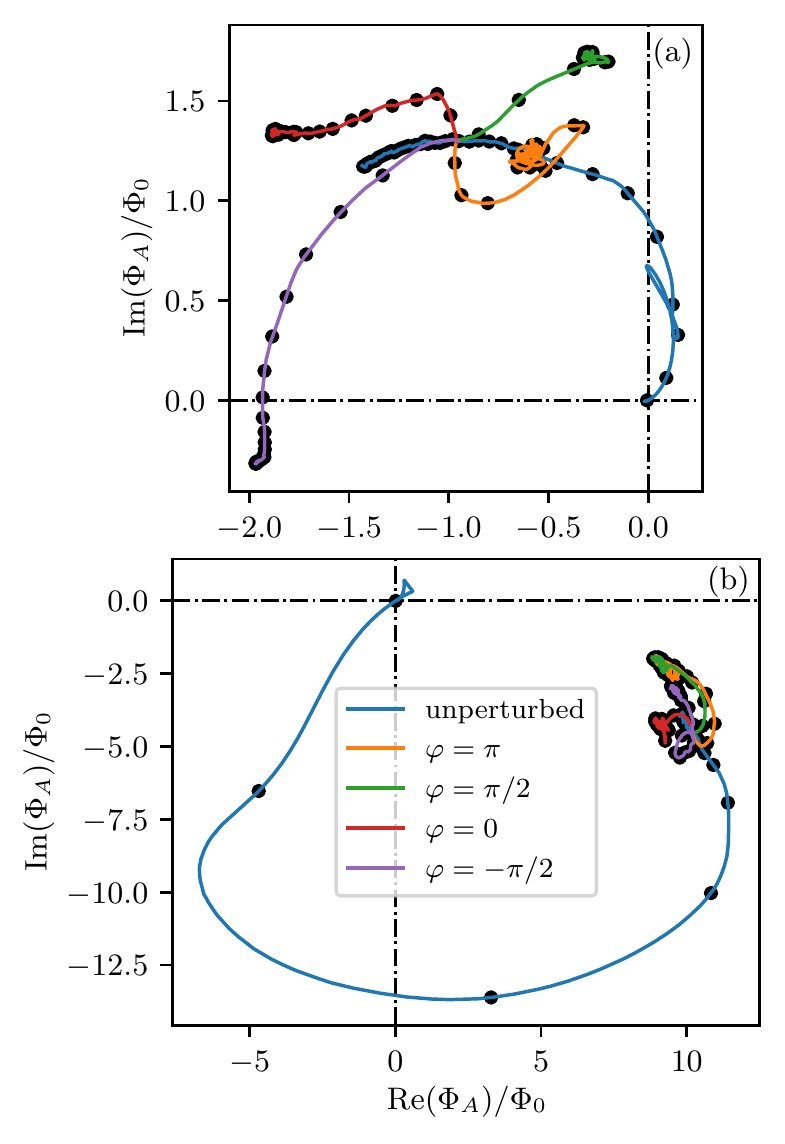}
\caption{The complex amplitude $\Phi_A (z)$ of the wakefield driven by the first bunch (a) and first ten bunches (b).
The blue lines correspond to the unperturbed beam.
The other colors correspond to perturbations with amplitude $A=0.4\Phi_0$ and different phases $\varphi$, which switch on at $z = 10^4 c/\omega_p$.
Black points follow at intervals of $c\tau_0$.}
\label{fig4-actinia}
\end{figure}

\begin{figure}[tb]
\includegraphics{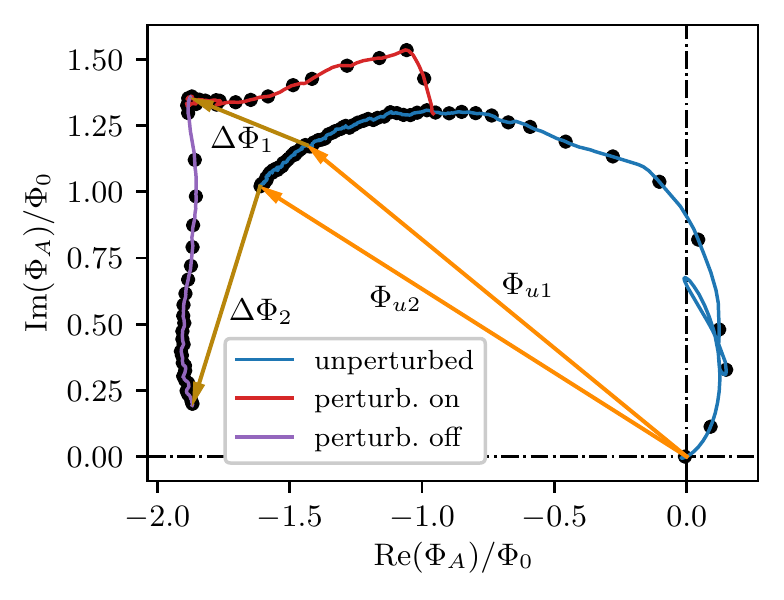}
\caption{Illustration of changes in the complex amplitude by the example of a single bunch and a perturbation with $A=0.4\Phi_0$ and $\varphi = 0$. }
\label{fig5-measure}
\end{figure}

After the perturbation is introduced, the wakefield generated by the beam first changes quickly and then stabilizes near a new equilibrium state, which depends on the phase and amplitude of the perturbation (figure~\ref{fig4-actinia}). This results in a non-zero ratio
\begin{equation}\label{e12}
    \delta \Phi = \frac{|\Delta\Phi|}{|\Phi_u|},
\end{equation}
where $\Phi_u$ is the complex wakefield amplitude with no perturbation, and $\Delta\Phi$ is the difference of complex amplitudes between the perturbed and unperturbed cases; both amplitudes are taken at the same $z$ to minimize the contribution of beam evolution due to other factors.
To quantify the effect of the perturbation on the beam, we measure $\delta \Phi$ at $z = z_1 \approx 42 c \tau_0$ and denote the corresponding values by the subscript ``1'' (figure~\ref{fig5-measure}).
In principle, the beam may not change much after the perturbation is switched on, but be destroyed after the perturbation is switched off.
To account for this possibility, we remove the perturbation at $z = z_1$ and analyze $\delta \Phi$ at $z = z_2 \approx 71 c \tau_0$, denoting the corresponding values by the subscript ``2'' (figure~\ref{fig5-measure}).

\begin{figure*}[tb]
\includegraphics{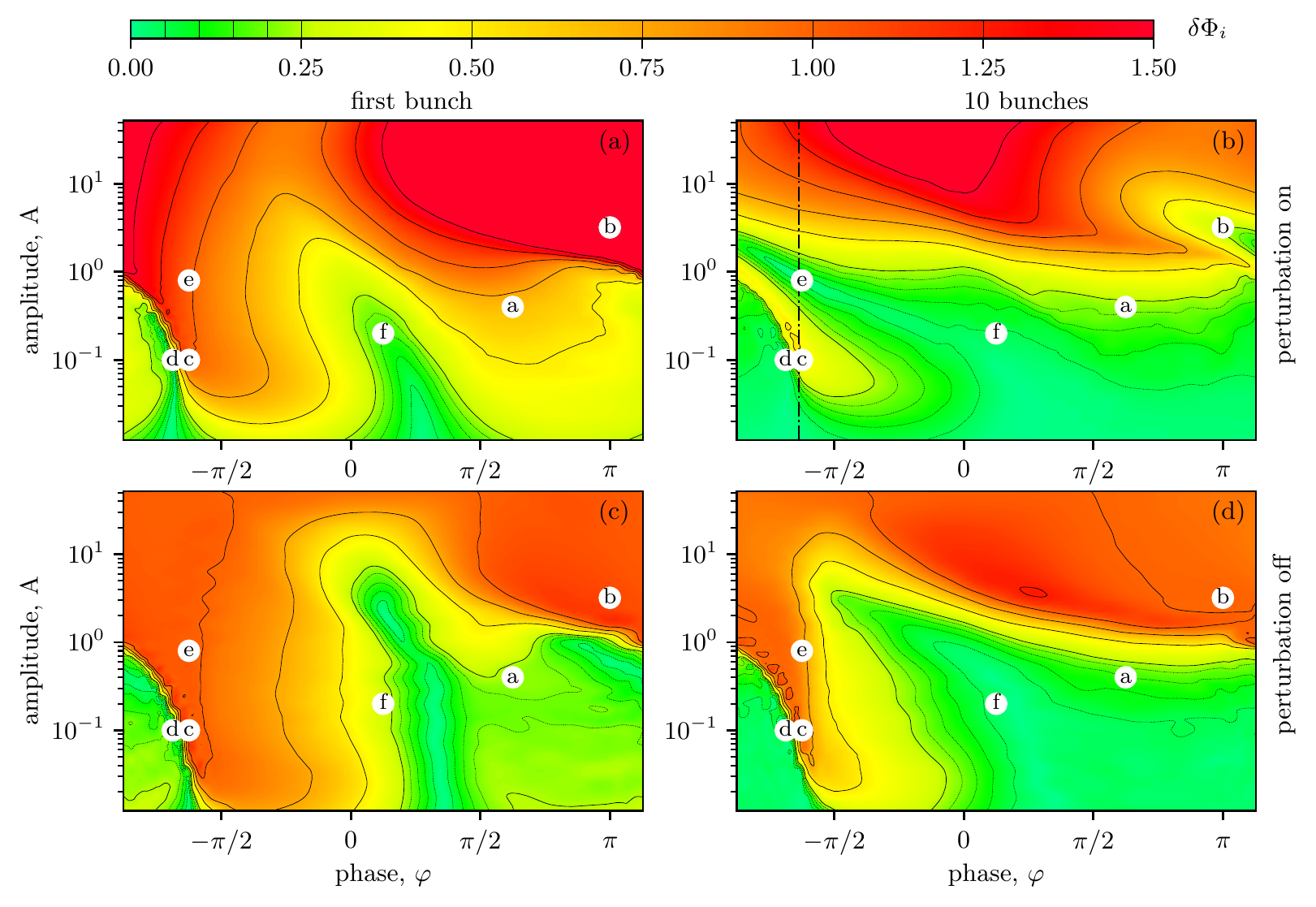}
\caption{Relative change of the complex amplitude $\delta \Phi_i$ ($i=1,2$) caused by perturbations of different amplitude $A$ and phase $\varphi$ in the cases of single bunch (a), (c) and train of ten bunches (b), (d) after switching the perturbation on (a), (b) and off (c), (d).
The points marked by letters in white circles correspond to the variants detailed in figure~\ref{fig7-details}. The dashed line in (b) is at $\varphi = -2$.}
\label{fig6-maps}
\end{figure*}

\section{Beam evolution}
\label{s3-effect}

Figure~\ref{fig6-maps} shows the effect of various perturbations on a single bunch and a train of ten bunches.
The relative change in complex amplitude $|\delta \Phi_{i}| = 0.25$ is large enough, for example, to shift the wave phase by $0.25/(\pi/2) \approx 16$\% of the interval favorable for witness acceleration and focusing, or change the wave amplitude by 25\%.
Both changes are marginally acceptable for the acceleration of a quality witness, so only in areas colored in shades of green can the beam be said to withstand the perturbation.
A train of many bunches is much more resilient than a single bunch, and can withstand perturbations with amplitudes up to about $\Phi_0$.
However, there are particularly dangerous perturbation phases in which a very small perturbing wave can destroy the beam.

Let us consider the changes in the beam and its wakefield in more detail.
For this, we analyze several representative cases, indicated in figure~\ref{fig6-maps} by letters in white circles.
If $A \ll \Phi_0$, the perturbation is much weaker than the own wakefield of the beam and is able to modify its structure only near the front and rear parts of the bunches, where the focusing field of the beam changes sign.
Therefore, the key to understanding the figure~\ref{fig6-maps} lies in the sign of the radial perturbation force at the leading fronts of the bunches.

\begin{figure*}[tb]
\includegraphics{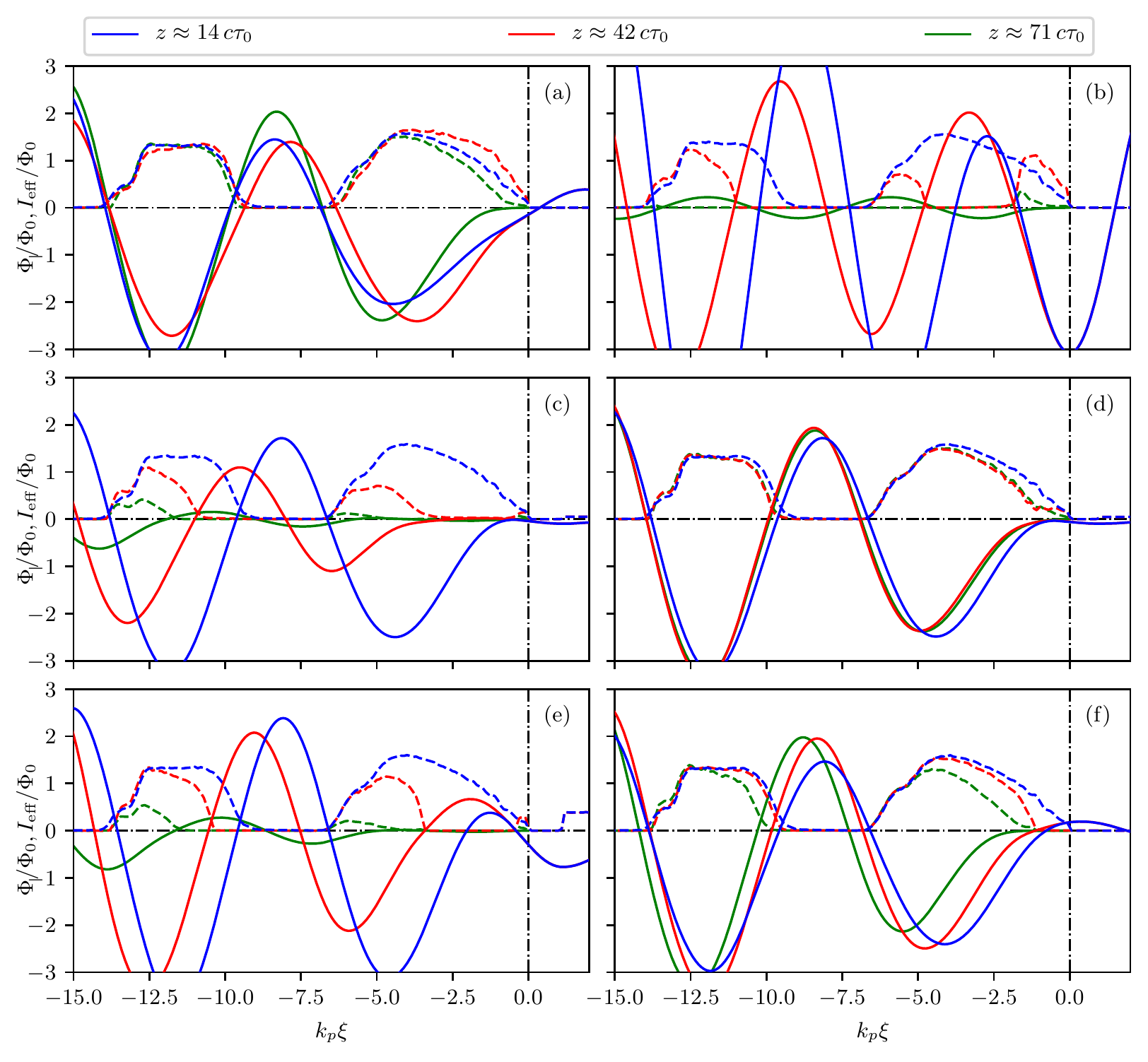}
\caption{The effective current $I_\text{eff}$ (dashed lines) and the wakefield potential $\Phi$ (solid lines) at different times (shown in the legend) for the variants marked in figure~\ref{fig6-maps}. The snapshots at $z \approx 14 c\tau_0$ (blue lines) are taken when the perturbation have just switched on and the equilibrium bunches have not yet change shape. The snapshots at $z \approx 42 c\tau_0$ (red lines) are taken just before switching the perturbation off and show the bunches and fields modified by the perturbation. The snapshots at $z \approx 71 c\tau_0$ (green lines) show the bunches and fields long after the perturbation is switched off.}
\label{fig7-details}
\end{figure*}

At the point $\xi=0$, the focusing force of the perturbing wave changes sign at $\varphi = \pm \pi/2$.
The current of the equilibrium beam does not increase immediately at $\xi=0$ (figure~\ref{fig3-perturbation}), so qualitative changes in the beam response occur at slightly lower values of the phase $\varphi$, at which the zero of the perturbation focusing force is at small negative $\xi$.

The phases $\varphi \gtrsim \pi/2$ and $\varphi \lesssim -\pi/2$ in figure~\ref{fig6-maps} correspond to perturbations which focus the very beginning of the beam.
The perturbation improves focusing of the first bunch, and its effective current increases, especially in the leading part that was weakly focused (figure~\ref{fig7-details}(a)).
This modifies the wakefield of the first bunch (point `a' in figure~\ref{fig6-maps}(a)) and moves field zeros slightly forward, because the ``center of mass'' of the first bunch shifts forward.
The subsequent bunches remain in the focusing phases, their contributions to the wave vary insignificantly, and $\delta \Phi_1$ becomes smaller as the share of the first bunch in the total wakefield decreases (figure~\ref{fig6-maps}(b)).
Similarly, when the perturbation is switched off, the wakefield of the first bunch changes a little (figure~\ref{fig6-maps}(c)), but not the wakefield of the bunch train (figure~\ref{fig6-maps}(d)).

At $\varphi \approx \pi$ and high perturbation amplitudes, a curious phenomenon is possible (figure~\ref{fig7-details}(b)).
The perturbing wakefield is stronger than the wakefield of the first bunch and heavily destroys this bunch (point `b' in figure~\ref{fig6-maps}(a)).
However, the perturbation is in phase with the wakefield of other bunches, so these bunches changes less (figure~\ref{fig6-maps}(b)).
It looks as if the perturbation wakefield replaces the wakefield of the first bunch.
Damage is caused by switching the perturbation off, which results in a strong phase change and partial destruction of all bunches in the train (figure~\ref{fig6-maps}(d)).

At $\varphi \approx -2$, the defocusing region of the perturbing wave overlaps the head of the first bunch and initiates a domino effect (figure~\ref{fig7-details}(c)).
The perturbation defocuses a bunch slice, this slice does not drive the wakefield necessary to keep the next slice focused, the next slice diverges because its transverse pressure is no longer balanced by the focusing force, its contribution to the wakefield disappears, then the next slice diverges, and so on.
The erosion of the first bunch causes the wakefield phase to move backward, which initiates the self-sustaining erosion of subsequent bunches.
As a result, even a long bunch train can be heavily destroyed by a perturbation of very small amplitude (point `c' in figures~\ref{fig6-maps}(a) and (b)).

The transition between the strong beam destruction by the domino effect and almost no beam change is very sharp (points `c' and `d' in figure~\ref{fig6-maps}(a) and (b)).
If the perturbation focuses the very beginning of the first bunch, then this bunch fully survives even if its body is in the defocusing phase of the perturbing wave (figure~\ref{fig7-details}(d)).
The own focusing field of the bunch overcomes the defocusing force of the perturbation.
The rest of the beam survives, too.
The stronger the perturbation, the more of the first bunch must be focused by the perturbation for the whole bunch to survive, the closer the transition to $\varphi = -\pi$  (figure~\ref{fig6-maps}).

It is interesting to see how the interaction regimes replace each other as the amplitude grows at \mbox{$\varphi \approx -2$} (the dashed vertical line in figure~\ref{fig6-maps}(b)).
At this phase, the leading edge of the first bunch is focused, followed by the defocusing half-period of the perturbation (figure~\ref{fig7-details}(e)).
At very small amplitudes, the defocusing force of the perturbation cannot exceed the own focusing force of the first bunch, and all bunches of the train change insignificantly.
At a higher amplitude, the domino effect destroys the beam (figure~\ref{fig7-details}(c)).
At an even higher amplitude, the first bunch is destroyed (point `e' in figure~\ref{fig6-maps}(a)), but its wakefield is replaced by that of the perturbation, and the other bunches of the train change little (figure~\ref{fig7-details}(e) and figure~\ref{fig6-maps}(b)), but collapse when the perturbation disappears (figure~\ref{fig6-maps}(d)).
Finally, at very large amplitudes, the entire beam is destroyed by the perturbation (figure~\ref{fig6-maps}(b)).

At phases $-\pi/2 < \varphi < \pi/2$, the beam head is defocused, the first bunch is partially destroyed, but the destruction wave does not propagate deep into the beam, as a considerable part of the first bunch falls into the focusing phase of the perturbation, survives and creates a wakefield of nearly the same phase as that without the perturbation (figure~\ref{fig7-details}(f) and points `f' in figure~\ref{fig6-maps}).

Summarizing the considered variants, a single bunch is easily destroyed by perturbations of rather low amplitude, except for certain perturbation phases at which the bunch fits the focusing phase of the perturbation and remains almost intact (figure~\ref{fig6-maps}(a)).
A train of many bunches, in turn, is resilient to perturbations of amplitude $A \lesssim \Phi_0$, except for perturbations which defocus the leading edge of the beam and initiate the domino effect (figure~\ref{fig6-maps}(b)).

\begin{figure}[tb]
\includegraphics{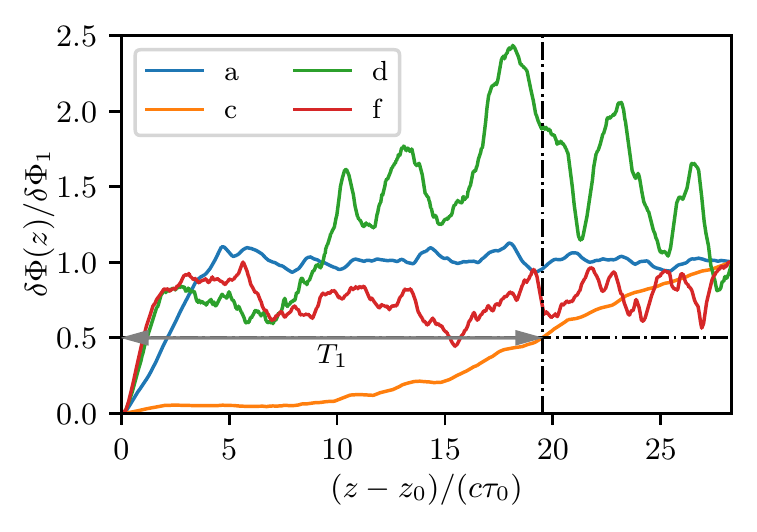}
\caption{Illustration of the beam response time to the perturbation: how the relative difference of complex amplitudes $\delta \Phi$ approaches $\delta \Phi_1$ for the train of ten bunches in several representative cases marked in figure~\ref{fig6-maps}(a) with the corresponding letters. The arrow shows the time $T_1$ for variant `c'.}
\label{fig8-tdef}
\end{figure}

\section{Beam response time}
\label{s4-time}

Under the perturbation, the beam changes in a complex way and approaches a new equilibrium state by no means exponentially.
The question arises how to characterize the duration of this process.
For this, we examine how the relative difference of complex amplitudes $\delta \Phi$ approaches the new equilibrium value.
When the perturbation is switched on, $\delta \Phi$ changes from zero to $\delta \Phi_1$  (figure~\ref{fig8-tdef}). When switched off, from $\delta \Phi_1$ to $\delta \Phi_2$.
The time $T_i$ ($i=1,2$) when $\delta \Phi$ passes halfway is the characteristic time of the process.

\begin{figure*}[tb]
\includegraphics{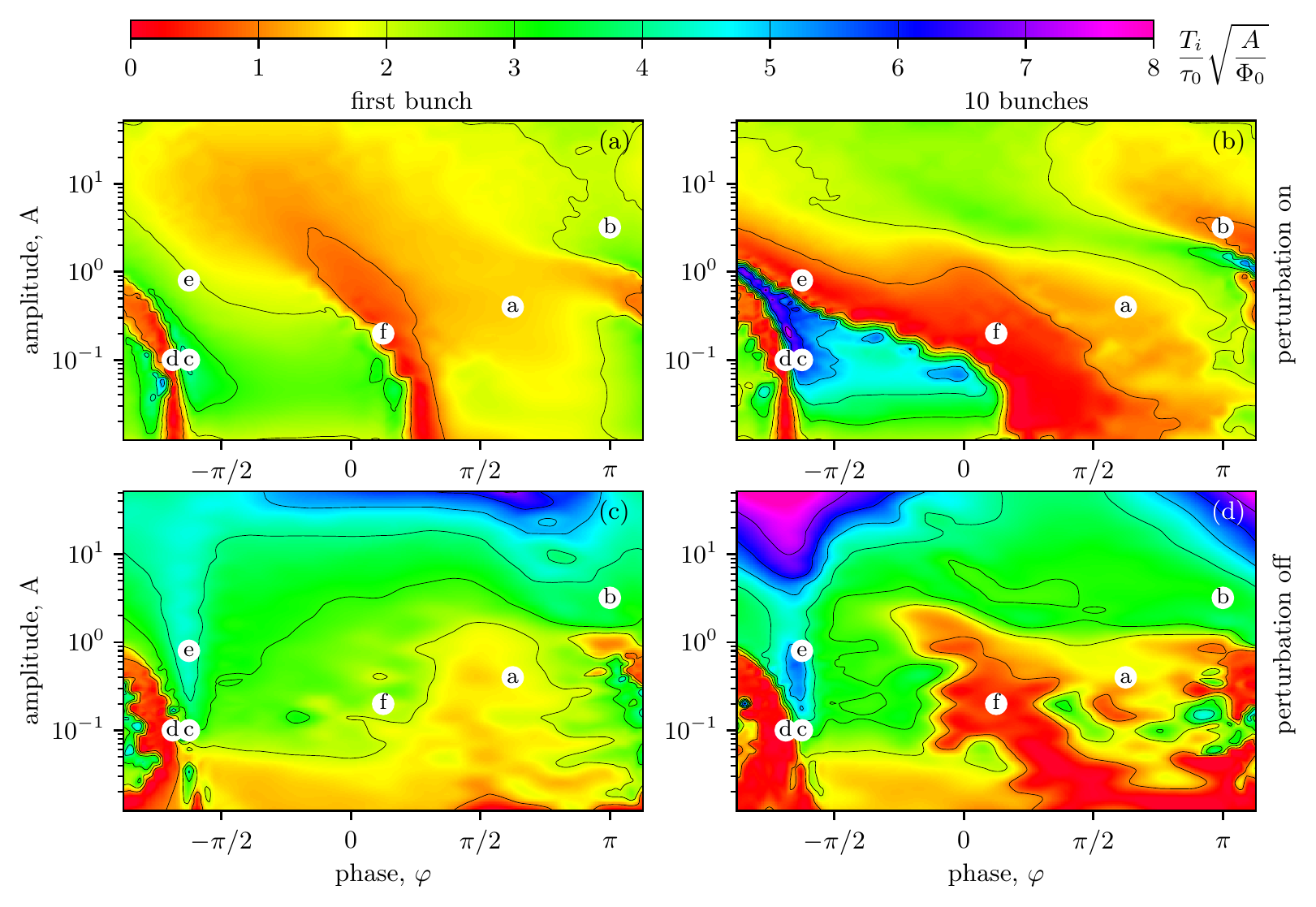}
\caption{Response time $T_i$ ($i=1,2$) of single bunch (a), (c) and train of ten bunches (b), (d) to switching the perturbation on (a), (b) and off (c), (d). The points marked by letters in white circles correspond to the variants detailed in figure~\ref{fig7-details}.}
\label{fig9-t}
\end{figure*}

In most cases, the beam response time, normalized to $\tau_0$, equals the inverse square root of the normalized perturbation amplitude, $(A/\Phi_0)^{-1/2}$, multiplied by a numerical factor of the order of unity (figure~\ref{fig9-t}).
This holds true for both single bunch and bunch train.
The only exception is the domino effect, which takes longer to develop in the bunch train (blue region in figure~\ref{fig9-t}(b)).
The red areas in figure~\ref{fig9-t}, which formally show a faster response, correspond to almost unchanging beams and represent the timescale of small field fluctuations (lines `d' and `f' in figure~\ref{fig8-tdef}).

\section{Summary}
\label{s5-summary}

An ultrarelativistic beam of charged particles in a plasma quickly comes to radial equilibrium with the wakefield.
When an external perturbation occurs in the plasma, created for example by another beam traveling ahead, the beam transforms into a different equilibrium state.
This new equilibrium state may or may not be close to the initial state, depending on the amplitude and phase of the perturbation.
If it is close, we say that the beam can resist the perturbation.
If not, the transition to another equilibrium state results in a partial loss of the beam charge, and we say that the beam is destroyed by the perturbation.
When the perturbing wakefield is removed, the beam does not return to its initial equilibrium state, but to another state, and this transition may be also accompanied by a loss of particles.

The typical perturbation amplitude, above which a train of many short bunches is strongly destroyed, equals the amplitude of the wakefield generated by a steep leading edge of the beam.
In our study, we took it as the natural unit of wakefield strength ($\Phi_0$).
It is approximately equal (within a factor of the order of unity) to the wakefield of a single short bunch of the ``resonant'' length (about a quarter of the plasma period).
A single bunch is less resilient than a train of many bunches and substantially changes its shape and wakefield under weaker perturbations.
However, if the perturbation is phased so that it defocuses the leading edge of the beam, even a long bunch train can be destroyed by a very weak field.
In this case, the bunches degrade one by one, similarly to the domino effect, when the destruction of one bunch initiates the destruction of the next bunch, and so on.

The typical transition from the unperturbed to perturbed equilibrium scales as the inverse square root of the perturbation amplitude and for an amplitude about $\Phi_0$ is of the order of the inverse betatron frequency of radial oscillations of the beam particles (\ref{e2}). The only exception is destruction of the bunch train by the domino effect, which takes longer.

\ack

This work was supported by the Russian Science Foundation, Project 20-12-00062. Simulations were performed on HPC cluster `Akademik V M Matrosov'  \cite{matrosov}.

\end{document}